# Curse of Dimensionality in Bayesian Model Updating

**Binbin LI[1], Zihan LIAO[2]**

[1]Assistant professor, ZJU-UIUC Institute, Zhejiang University, China.
E-mail: bbl@zju.edu.cn

[2]Ph.D. Student, College of Civil Engineering and Architecture, Zhejiang University, China.
E-mail: zihanliao@intl.zju.edu.cn

**ABSTRACT**

Bayesian approach provides a coherent framework to address the model updating problem in structural health monitoring. The current practice, however, only focuses on low-dimension model (generally no more than 20 parameters), which limits the accuracy and predictability of the updated model. This paper aims at understanding the curse of dimensionality in Bayesian model updating, and thus proposing feasible strategies to overcome it. An analytical investigation is conducted, which allows us to answer fundamental questions in Bayesian analysis, e.g., where the posterior mass locates and how large of it comparing to the prior volume. The key concept here is the distance from the prior to the posterior, which makes the parameter estimation really difficult in high-dimension problems. In this sense, not only the dimensionality matters, but also the multi-modality, the pronounced degeneracy, and other factors that influence the prior-posterior distance matter.

**KEYWORDS:** *model updating, Bayesian inference, dimensionality, high-probability set, subset simulation*

## 1. GENERAL INSTRUCTIONS

The core of Bayesian analysis is to compute of evidence and posterior given the model and data. One can expand the joint probability of a $d$ dimensional parameter vector $\boldsymbol{\theta} \in \Omega$ (in a given model $\mathcal{M}$) and the data $\boldsymbol{D}$ in two different ways:

$$
\begin{array}{ccccccc}
\Pr(\boldsymbol{D}|\boldsymbol{\theta},\mathcal{M}) & \Pr(\boldsymbol{\theta}|\mathcal{M}) & = \Pr(\boldsymbol{\theta},\boldsymbol{D}|\mathcal{M}) = & \Pr(\boldsymbol{D}|\mathcal{M}) & \Pr(\boldsymbol{\theta}|\boldsymbol{D},\mathcal{M}) \\
\mathcal{L}(\boldsymbol{\theta}) & \times \;\; \pi_{\boldsymbol{\theta}}(\boldsymbol{\theta})d\boldsymbol{\theta} & = \;\;\cdots\;\; = & z & \times \;\; p_{\boldsymbol{\theta}}(\boldsymbol{\theta})d\boldsymbol{\theta} \\
Likelihood & \times \;\; Prior & = \;\; Joint \;\; = & Evidence & \times \;\; Posterior \\
& Inputs & \Rightarrow \;\;\cdots\;\; \Rightarrow & & Outputs
\end{array}
\tag{1}
$$

which directly yields the Bayes' theorem and the evidence, respectively, as

$$p_{\boldsymbol{\theta}}(\boldsymbol{\theta}) = \frac{1}{z} \mathcal{L}(\boldsymbol{\theta})\pi_{\boldsymbol{\theta}}(\boldsymbol{\theta}) \tag{2}$$

$$z = \int_\Omega \mathcal{L}(\boldsymbol{\theta})\pi_{\boldsymbol{\theta}}(\boldsymbol{\theta})d\boldsymbol{\theta} \tag{3}$$

The inputs to computation are the prior $d\Pi_{\boldsymbol{\theta}}(\boldsymbol{\theta}) = \pi_{\boldsymbol{\theta}}(\boldsymbol{\theta})d\boldsymbol{\theta}$ and the likelihood $\mathcal{L}(\boldsymbol{\theta})$, whilst the desired outputs are the evidence $z$ and the posterior $dP_{\boldsymbol{\theta}}(\boldsymbol{\theta}) = p_{\boldsymbol{\theta}}(\boldsymbol{\theta})d\boldsymbol{\theta}$. Here, we emphasize the probability mass function (PMF) (e.g., $\Pi_{\boldsymbol{\theta}}(\boldsymbol{\theta})$ and $P_{\boldsymbol{\theta}}(\boldsymbol{\theta})$), which denotes an accumulated amount of probability. The probability density function (PDF) (e.g., $\pi_{\boldsymbol{\theta}}(\boldsymbol{\theta})$ and $p_{\boldsymbol{\theta}}(\boldsymbol{\theta})$) is its differential $d(mass)/d(volume\ of\ \boldsymbol{\Theta})$. Analytical calculation of the posterior PDF and evidence (Eqs. (2) and (3)) is generally infeasible, so that one has to rely on the approximation methods, e.g., the Laplace approximation, the variational method, and the Monte Carlo sampling (MCS). The MCS approach, including the Markov Chain Monte Carlo (MCMC) sampling, provides a universal way for Bayesian inference. However, it is known that many MCS algorithms, e.g., importance sampling (Au and Beck, 2003) and Metropolis-Hastings algorithm (Betancourt, 2017), does not scale well with the dimension of parameter. That is their sampling efficiency reduces with the increase of the number of parameters. It is known as the "curse of dimensionality" in the statistical community. Rather than proposing an algorithm scaling well with dimension, this paper aims at understanding the curse of dimensionality by answering the following fundamental questions:

"Where does the posterior mass locate in the posterior distribution?"

"How large is the posterior space comparing to the volume of its prior?"

"What is the effective dimension of the problem?"

The above questions characterize the Bayesian inference problem in high dimension. Knowing them not only allows us to appreciate the difficulty in high-dimensional inference, but also provides insights on how to overcome the curse of dimensionality.

This paper starts with an interpretation of evidence from the perspective of the complementary cumulative distribution function (CCDF) of likelihood $\mathcal{L}(\boldsymbol{\theta})$ given $\boldsymbol{\theta} \sim \pi_{\boldsymbol{\theta}}(\boldsymbol{\theta})$. It lays the theoretical foundation of the paper, because it transforms a $d$-dimensional problem into a one-dimensional analysis. Answers for the above questions are then provided based on information theory, and the situation of high dimensional inference is also analyzed.

## 2. CCDF REPRESENTATION

The evidence in Eq. 3 is represented in terms of a $d$-dimensional integration. With the aid of CCDF, one can write equivalently (Handley and Lemos, 2019),

$$z = \int_0^{l_{sup}} X(l) dl = \int_0^1 L(x) dx \tag{4}$$

where $X(l) = \Pr(\mathcal{L}(\boldsymbol{\theta}) > l) \in [0,1]$ given $\boldsymbol{\theta} \sim \pi_{\boldsymbol{\theta}}(\boldsymbol{\theta})$, i.e., the complementary cumulative distribution function (CCDF) of likelihood $\mathcal{L}(\boldsymbol{\theta})$, and $l_{sup}$ denotes the supreme value of the likelihood function. $L(x)$ is the inverse function of $X(l)$, and can be calculated as $L(x) = \sup\{l: X(l) > x\}$. The CCDF $X(l)$ is a non-increasing function, so its inverse function $L(x)$. Therefore, $L(x)$ is named as "sorted likelihood function" (SLF) in this paper. Intuitively, $X(l)$ accumulates the prior mass of $\boldsymbol{\theta}$ where likelihood values are greater than $l$, while $L(x)$ gives the value $l$ such that $x$ is the fraction of prior draws with likelihood values larger than $l$.

In the Bayesian approach, the unknown parameter $\boldsymbol{\theta} \in \mathbb{R}^d$ is regarded as a random variable (RV), so are the likelihood function $\mathcal{L}(\boldsymbol{\theta})$, the CCDF $X(\mathcal{L})$, and the SLF $L(X)$. Note that there is a one-to-one correspondence between $X$ and $\boldsymbol{\theta}$, though they have different dimensions. Interestingly, RV $X$ has a uniform distribution $U(0,1)$ not matter the prior PDF $\pi_{\boldsymbol{\theta}}(\boldsymbol{\theta})$ is, because it is a CCDF. Since both $\mathcal{L}(\boldsymbol{\theta})$ and $L(X)$ represent the same likelihood although in different parameterizations, they should follow the same probability distribution. In fact, one can show that their cumulative distribution functions (CDFs) are identical: $F_L(l) = \Pr(L(X) \leq l) = \Pr(X(L) \geq X(l)) = 1 - X(l) = \Pr(\mathcal{L}(\boldsymbol{\theta}) < l) = F_{\mathcal{L}}(l)$. The relations of RVs $\boldsymbol{\theta}$, $\mathcal{L}$, $X$, and $L$ are summarized in Figure 1. According to the above analysis, one can obtain the posterior PDF of $X$ from $\boldsymbol{\theta} \sim p_{\boldsymbol{\theta}}(\boldsymbol{\theta})$ shown in Eq. 2 as

$$p_X(x) = \frac{1}{z} L(x) \tag{5}$$

Since a valid PDF must be normalized to have a unit mass, one can further verify Eq. 4 by taking an integration on both sides of Eq. 5.

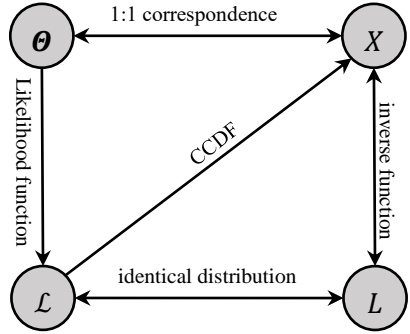

**Figure 1.** Relations of random variables in Bayesian inference.

For better understanding the above concepts, we take a simple example for illustration. Assume that the unknown parameter $\boldsymbol{\theta} \in \mathbb{R}^d$ has a uniform prior within a unit sphere, i.e., $\pi_{\boldsymbol{\theta}}(\boldsymbol{\theta}) = \Gamma(d/2 + 1)\pi^{-d/2} \mathbb{1}[\boldsymbol{\theta}^T \boldsymbol{\theta} \leq 1]$, and the likelihood is given by a multivariate normal $\mathcal{L}(\boldsymbol{\theta}) = \exp(-0.5\sigma^{-2}\boldsymbol{\theta}^T\boldsymbol{\theta})$, where $\sigma \ll 1$ so that almost all the likelihood is well contained by the prior domain. It then yields a multivariate normal posterior with the following PDF

$$p_{\boldsymbol{\theta}}(\boldsymbol{\theta}) = \frac{1}{z}\mathcal{L}(\boldsymbol{\theta})\pi_{\boldsymbol{\theta}}(\boldsymbol{\theta}) = (2\pi\sigma^2)^{-d/2} \exp\left(-\frac{1}{2\sigma^2}\boldsymbol{\theta}^T\boldsymbol{\theta}\right) \tag{6}$$

with $z = \Gamma(d/2 + 1)(2\sigma^2)^{d/2}$ discarding the tails outside the domain. The CCDF and SLF can be calculated respectively as $X(l) = (-2\sigma^2 \ln l)^{d/2}$ and $L(x) = \exp(-0.5\sigma^{-2}x^{2/d})$, yielding the posterior PDF of $X$ as

$$p_X(x) = \frac{1}{z} L(x) = \frac{1}{\Gamma(d/2 + 1)(2\sigma^2)^{d/2}} \exp\left(-\frac{1}{2\sigma^2}x^{2/d}\right) \tag{7}$$

With the introduction of CCDF $X(l)$, the Bayesian inference for $\boldsymbol{\theta}$ in a $d$ dimension can be equivalently expressed as an inference for $X$ in a single dimension, thus greatly releasing the burden of analysis.

## 3. BAYESIAN INFERENCE IN A HIGH DIMENSION

In order to answer the questions proposed in the Introduction section, we first give the definition of "high probability set".

**Definition**: For a small real number $\varepsilon > 0$, the high probability set (HPS) $A_\varepsilon$ with respect to $p_X(x)$ is $A_\varepsilon = \{x \in [0,1]: |\ln p_X(x) - G| \leq \varepsilon\}$, where $p_X(x)$ is the posterior PDF of CCDF $X$ and $G$ is the information gain from data $\boldsymbol{D}$ (or Kullback–Leibler distance from prior to posterior), defined as $G = \int_0^1 p_X(x) \ln p_X(x)\, dx$ because CCDF $X$ has a uniform prior.

From the definition of the information gain, it is the expectation of $\ln p_X(x)$ given $X \sim p_X(x)$. Therefore, the HPS $A_\varepsilon$ defines an area around the posterior mean of $\ln p_X(x)$. For HPS $A_\varepsilon$, we state the following theorem without proof.

**Theorem**: the HPS $A_\varepsilon$ has the following properties:

(1) $\Pr(A_\varepsilon) \geq 1 - \sigma_p^2/\varepsilon^2$, where $\sigma_p^2 = \mathrm{E}\{[\ln p_X(x) - G]^2\}$, i.e., posterior variance of $\ln p_X(x)$;

(2) $(1 - \sigma_p^2/\varepsilon^2)e^{-(G+\varepsilon)} \leq \mathrm{Vol}(A_\varepsilon) \leq e^{-G+\varepsilon}$, where $\mathrm{Vol}(A_\varepsilon)$ represents the volume of $A_\varepsilon$.

The first property indicates most probability mass focuses within this area by properly selecting $\varepsilon$, e.g. $\Pr(A_\varepsilon) > 0.89$ if $\varepsilon = 3\sigma_p$; while the second property shows the volume of the high probability set is close to $e^{-G}$ of the prior which has a unit volume. Since $G$ can be the order of thousands, the high probability set conquers an extremely small volume of prior space, but, astonishingly, it contains most probability mass. Now, the answers to the first two questions should be clear: most of posterior mass locates within HPS $A_\varepsilon$ which centers around the information gain $G$; the posterior occupies a small fraction $e^{-G}$ of the prior volume. The information gain $G$ quantifies the probabilistic distance from the prior to the posterior as well as the volume of the posterior relative to that of the prior.

As for the "effective" dimension of the Bayesian inference, we adopt the Bayesian model dimensionality proposed in (Gelman *et al.*, 2014):

$$d_e = 2\sigma_p^2 = 2\mathrm{E}\left\{[\ln p_X(x) - G]^2\right\} \tag{8}$$

which corresponds to the dimension of parameters whose posterior is a multivariate normal distribution (as illustrated later). Based on the above analysis, one then can construct the relation between the width of HPS $A_\varepsilon$ and the effective dimension. That is the width of $A_\varepsilon$ is proportional to $\sqrt{d_e}$. For example, one can easily find $\Pr(A_\varepsilon) \geq 0.5$ by setting $\varepsilon = \sqrt{d_e}$

The determination of the HPS $A_\varepsilon$ and the effective dimension requires the posterior distribution of $\ln p_X(x)$, and its mean and variance. From Eq. 5, one has $\ln p_X(x) = \ln L(x) - \ln z$. Therefore, one can also first find the posterior distribution of $Y = \ln L(x)$, and then obtain $G = \mathrm{E}[Y] - \ln z$ and $\sigma_p^2 = \sigma_Y^2$, which is the posterior variance of RV $Y$. Suppose we are able to compute the CCDF $X(l)$, e.g. using the subset simulation (Betz *et al.*, 2018), the posterior PDF of $Y$ is given by

$$p_Y(y) = p_X(x(y)) \left|\frac{dx}{dy}\right| = \frac{e^y}{Z}\frac{dX(e^y)}{dy} \tag{9}$$

Consider again the example in last section. Substituting $X(l) = (-2\sigma^2 \ln l)^{d/2}$ into Eq. 9 gives

$$p_Y(y) = \frac{1}{\Gamma(d/2)} \exp(y)\, (-y)^{d/2-1} \tag{10}$$

which can be calculated to have mean of $-d/2$ and variance of $d/2$. We then have the information gain $G = -d/2 - d/2 \ln(2\sigma^2) - \ln[\Gamma(d/2+1)] \approx -d/2 \ln(\sigma^2 d)$ according to the Stirling's formula and the effective dimension $d_e = 2\sigma_p^2 = d$, i.e., it recovers the true dimension of the problem. This should not be a surprise, because Bayesian model dimensionality is defined to be so. An illustration of the sorted likelihood and $p_Y(y)$ is provided in Figure 2 for $\sigma = 0.01$ and $d = 100$. The HPS $A_\varepsilon$ in terms of $\ln L$ is shown by the red line enclosed by mean $\pm$ 3 standard deviation. Interestingly, the HPS $A_\varepsilon$ does not include the posterior mode of the unknown parameter $\boldsymbol{\theta}$ corresponding to $\ln L = 0$.

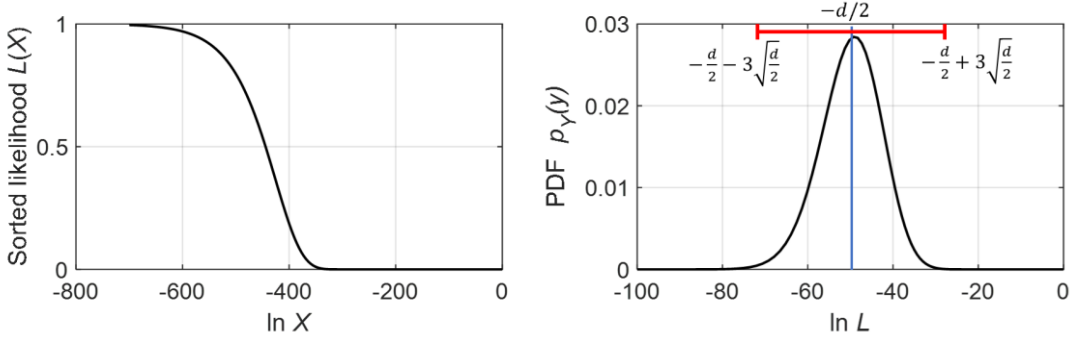

**Figure 2.** SLF $L(x)$ and posterior PDF $p_Y(y)$

Note: HPS $A_\varepsilon$ is indicated by the interval of mean $\pm$ 3 standard deviation.

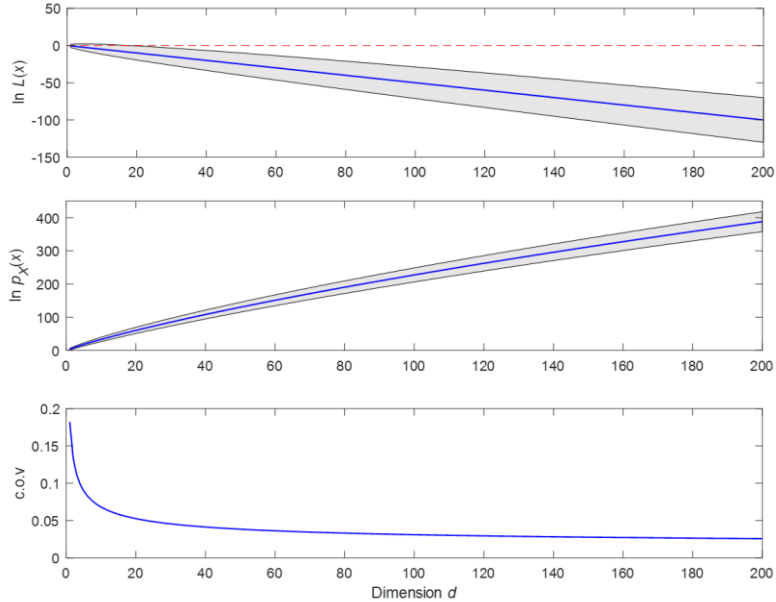

**Figure 3.** Behavior of HPS $A_\varepsilon$ changes with dimension $d$.

Note: HPS $A_\varepsilon$ is indicated by shaded area.

For further illustration, the plot of HPS $A_\varepsilon$ changing with dimension $d$ is provided in Figure 3 by fixing $\sigma = 0.01$. As can be seen, HPS $A_\varepsilon$ is farther and farther away from $\ln L = 0$ as the dimension increases, so that obtaining a sample around the mode of $\boldsymbol{\theta}$ becomes a rare event, which is a prominent difference with the low dimensional problems. The information gain $G$, i.e., the distance from the prior to the posterior, increases almost linearly with dimension $d$, which means the volume of HPS $A_\varepsilon$ decreases exponentially; therefore, it becomes harder and harder to locate it from prior. In addition, although the standard deviation $\sigma_p \propto \sqrt{d}$ increases with $d$, the coefficient of variation (c.o.v.) $\propto 1/\sqrt{d}\ln(\sigma\sqrt{d})$, which is the ratio of standard deviation to the mean, decreases as the dimension becomes higher, indicating the width of high probability set shrinks in high dimensional space so that navigating within the high probability set is difficult as well.

**CONCLUSIONS**

Bayesian inference with high dimensional parameters is difficult. Understanding the behavior of the posterior distribution is critical for overcoming the curse of dimensionality. The high probability set (HPS) is defined in the paper from the perspective of CCDF. It centers around the information gain from data with a width quantified by the standard deviation of log posterior PDF. This width also gives an intuitive tool to define the "effective dimension" of the parameter space.

It shows that the HPS in high-dimensional parameter space is far away from the mode of the posterior distribution. Therefore, a rare number of posterior samples locates around the posterior mode. The information gain almost linearly increases with the dimension, indicating the distance from the prior to posterior is long for high-dimensional space. The posterior only conquers an exponentially small volume of the prior space.